%% file: cameraReady.tex
\documentclass[conference]{IEEEtran}

\usepackage{comment}
\usepackage{graphicx,subfigure,url,balance,comment} 
\usepackage{algorithmic} 
\usepackage{multirow} 
\usepackage{amsmath} 
\usepackage{amsfonts} 
\usepackage{epstopdf} 
\usepackage{caption2}
\usepackage{upquote}
\usepackage{tabularx}
\usepackage{authblk}
\usepackage{etoolbox}
\usepackage{color}
\usepackage[misc]{ifsym}
\usepackage[ruled]{algorithm2e}
\makeatletter
\patchcmd{\maketitle}{\@copyrightspace}{}{}{}
\makeatother

\usepackage{float}

\SetKwRepeat{doWhile}{do}{while}

\begin{document}

\title{The Internet of Responsibilities - Connecting Human Responsibilities using Big Data and Blockchain}

\author{
  Xuejiao Tang$^1$, Jiong Qiu$^2$(\Letter), Wenbin Zhang$^3$, Ibrahim Toure$^4$, Mingli Zhang$^5$\\
  
  Enza Messina$^6$, Xueping Xie$^7$, Xuebing Wang$^2$ and Sheng Yu$^2$ \\
  
  $^1$Leibniz University Hannover, Germany $^2$Hangzhou Quanshi Software Co., Ltd, China\\
  $^3$University of Maryland, Baltimore County, USA $^4$Doninya Inc., USA $^5$McGill University, Canada\\
      $^6$University of Milano-Bicocca, Italy $^7$Hangzhou Dianzi University, China\\
  xuejiao.tang@stud.uni-hannover.de, colin\_qiu@hotmail.com, wenbinzhang@umbc.edu,
  ibrahim.iba.toure@gmail.com \\ 
  mingli.zhang@mcgill.ca, enza.messina@unimib.it, xxp@hdu.edu.cn, \{wangxuebing1109, yusheng0129\}@qq.com
  }

\maketitle 
\input{abstract.tex}
\vspace{-1mm}
\input{introduction.tex}

\vspace{-1mm}
\input{design.tex}

\vspace{-1mm}
\input{experiment.tex}
\vspace{-1mm}
\input{conclusion.tex}

\nocite{*}
\bibliographystyle{abbrv}
\bibliography{ref}

\end{document}

%% file: abstract.tex
\begin{abstract}
\label{sect:abstract}
 Accountability in the workplace is critically important and remains a challenging problem, especially with respect to workplace safety management. In this paper, we introduce a novel notion, the Internet of Responsibilities, for accountability management. Our method sorts through the list of responsibilities with respect to hazardous positions. The positions are interconnected using directed acyclic graphs (DAGs) indicating the hierarchy of responsibilities in the organization. In addition, the system detects and collects responsibilities, and represents risk areas in terms of the positions of the responsibility nodes. Finally, an automatic reminder and assignment system is used to enforce a strict responsibility control without human intervention. Using blockchain technology, we further extend our system with the capability to store, recover and encrypt responsibility data. We show that through the application of the Internet of Responsibility network model driven by Big Data, enterprise and government agencies can attain a highly secured and safe workplace. Therefore, our model offers a combination of interconnected responsibilities, accountability, monitoring, and safety which is crucial for the protection of employees and the success of organizations.
\end{abstract}

%% file: introduction.tex
\section{Introduction}
\label{sect:intro}
Together with delegation and representation, accountability is one of the cornerstones of democracy. Delegation involves endowing another party with the discretion to act, representation is about the interests that are at stake, and accountability ensures that the exercise of discretion is checked. In the last few years, various patents pertaining to responsibility management have been filed and can be classified into the following schemes: responsibility data production (collection), responsibility process design, responsibility data exchange, responsibility data evaluation, responsibility clarity \cite{10.1093/jopart/mus034}. However, patents are limited in that they cannot describe the systematic and complete literature that encompasses the close-loop management of responsibility. Among them, Hollender et al. expanded the risk alarm management to reflect the responsibility of each link of the overall process~\cite{hollender2012method}. In addition, Brandsma et al. proposed accountability cube to quantify accountability \cite{10.1093/jopart/mus034}. Nonetheless, despite these efforts, a complete responsibility assignment management system is still lacking. On the other hand, data-driven AI is increasingly being used to improve every aspect of our lives \cite{zhang2019faht,zhang2017hybrid,zhang2016using}. In this paper, a big data driven accountability management model, the Internet of Responsibilities model (IoR), is therefore proposed to improve the quality and safeguard of enterprises and institutions. Accountability in a production environment is implemented through the IoR model which will promote the staffs' awareness of responsibility. Our model has been deployed by the Chinese governments and companies to establish the accountability mechanism. The results of the initial evaluation showed an obvious increase in responsibility scores and safety awareness.


%% file: design.tex
\section{System Design}
\label{sect:design}
\subsection{The Internet of Responsibilities System}
\begin{figure}[!h]
    \vspace{-5mm}
    \centering
    \includegraphics[width=\linewidth, height=0.25\textheight]{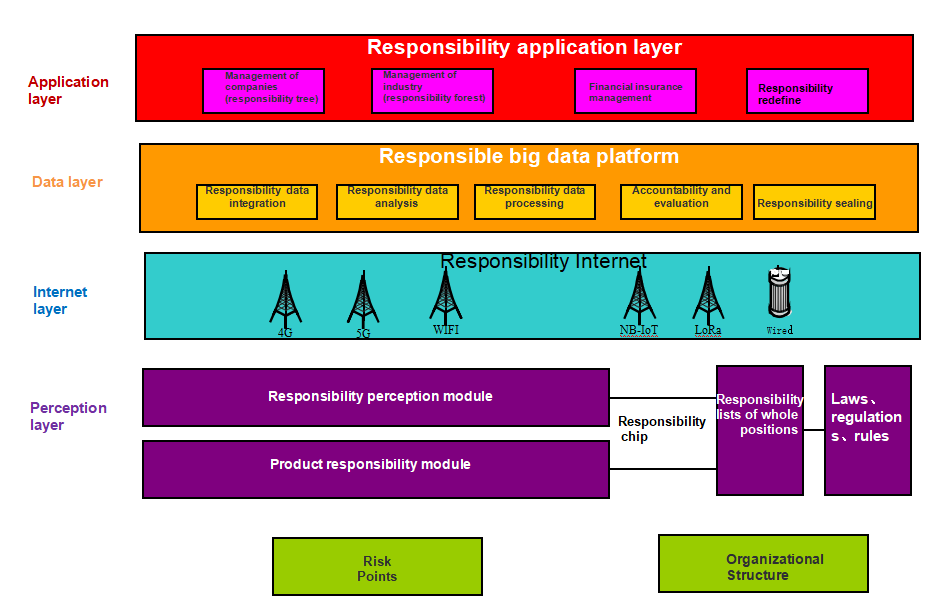}
    \caption{The Internet of Responsibility model.}
    \label{fig:figure1}
    \vspace{-4mm}
\end{figure}

As presented in Figure~\ref{fig:figure1}, IoR model consists of four layers: perception layer, network layer, data layer and application layer. These layers are detailed thereafter. 

\begin{enumerate}
\item{ \textit{Perception Layer}. This layer collects the responsibility data. Starting from the bottom, different types of responsibility data are collected from a list of hazardous locations. Our model takes into account hazardous locations and the hierarchy of the organization. The data collected is transformed into a unified data model through the perception module, using sophisticated algorithms, and the product module, based on hardware design. In addition, the reliability of hardware system and relative information on the perception layer is ensured by the responsibility chip, designed in a blockchain mode to record authentication information like device manufacturer authentication information, device process data, etc.
}

\item{\textit{Internet layer}. The network layer uploads the responsibility data collected at the perception layer to the Internet via Fiber, 4G, 5G, WiFi, NB-IoT, or LoRa.}

\item{\textit{Data layer.} This layer adopts semi-structured governance of the responsibility data, i.e., data integration, data analysis for all of the hazardous locations, data processing, accountability and evaluation (score), and encryption of responsibility data. By using directed acyclic graph (DAG), responsibility nodes with responsibility data uploaded to the \textit{Internet layer} from the \textit{perception layer}, are sorted according to the hierarchy of the organization and the risk level, which results in a responsibility tree for the company. For each hazardous position, all of the relevant information will be stored in the responsibility nodes, which is implemented on the blockchain.}

\item{\textit{Application layer.} The application layer allows different types of companies such as financial institutions, government agencies; to monitor, automatically detect abnormal events, and send alerts to the responsible parties. Furthermore, this layer is used to manage and reorganize the responsibility list of hazardous locations.} \newline
\end{enumerate} 

\subsection{Details of IoR model}
This section outlines the key technologies of IoR model which enable the implementation of various layers. 
\subsubsection{Responsibility module of Roles}
Multiple multivariate groups are defined to describe the responsibility lists of the positions:
\begin{itemize}
 \item[-] Definition 1: Role (roleid, rolename, $<$RList$>$, $<$PersonList$>$, $<$ObjectList$>$, $<$AlarmList$>$, remarks). Role refers to the position of each employee in the company. A position contains either one or more roles.
 \item[-] Definition 2: RList (rid, Rtype, Rname, $<$RdataList$>$, Rscore, Rrequirement, Rremarks). RList refers to responsibility list which consists of id, type, responsibility data, scoring items, basis (laws, regulations, rules) and remarks.
 \item[-] Definition 3: PersonList (personid, personname, $<$DatetimeList$>$, $<$SpaceList$>$, $<$RoleidList$>$, remarks). PersonList contains the staffs' information with a given role,
 DatetimeList defines the start and end time, SpaceList means the location distributions of each person. One person can have multiple roles.
\item[-] Definition 4: ObjectList (oid, objectname, $<$QuantityList$>$, remarks). ObjectList contains information about goals. 
\item[-] Definition 5: AlarmList (aid, aname, alevel, asource, $<$AlarmRNodeList$>$, remarks). AlarmList contains items from hazardous positions, indicating the id, name, risk level (A, B, C class are defined to indicate risk level decreasingly), risk source and defines AlarmRNodeList that can automatically generate responsibility lists over time.
\end{itemize}

\subsubsection{Hierarchy of Responsibilities in Organization}
\begin{figure}[!h]
	\vspace{-5mm}
	\centering
	\includegraphics[width=\linewidth, height=0.10\textheight]{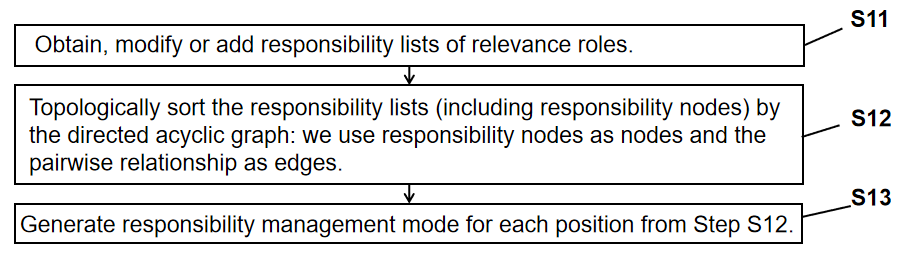}
	\caption{Directed acyclic graph for responsibility lists.}
	\label{fig:figure2}
	\vspace{-2mm}
\end{figure}

To indicate the hierarchy of responsibilities in the organization among responsibility lists, positions are interconnected using DAGs as shown in Figure~\ref{fig:figure2}.
\begin{itemize}
 \item[-] S11: Obtain, modify or add responsibility lists of relevance roles.
 \item[-] S12: Topologically sort the responsibility lists (including responsibility nodes) by the directed acyclic graph: we use responsibility nodes as nodes and the pairwise relationship as edges.
 \item[-] S13: Generate responsibility management mode for each position from Step S12.
\end{itemize}

\subsubsection{Automatic evaluation and management}
The responsibility score is calculated by automatic evaluation of completion score and execution score.

The completion score with a total of 1000 points is divided into two parts: daily work score and additional work score. The default proportion is set to 750 points for daily work score and 250 points for additional work score. These scores reflect the daily completion status of relative responsibilities. The proportion between the daily work score and additional work score can be reset. The scores are calculated based on the timely submission of required materials by staff members proving the completion of the responsibilities assigned.
The system computes the execution scores of each staff member with responsibilities using the results from materials submitted and co-workers' evaluations.
\subsubsection{Responsibility Perception Design}
Responsibility perception is designed as follows: 
\begin{itemize}
	\item[-] Step 1: Sort responsibility lists and information to responsibility sets, which contain responsibility nodes. Each responsibility set includes one or multiple starting nodes, process nodes, and terminal nodes.
	\item[-] Step 2: Correlate responsibility sets with product default status, which results in a status table of responsibilities. 
	\item[-] Step 3: Add the methodology describing how the status table of responsibilities generates \textit{responsibility data} (from step 2) into production.
	\item[-] Step 4: When production status changes, responsibility data gets generated.
	\item[-] Step 5: Responsibility score based on \textit{responsibility data} are computed while generating relative data.
\end{itemize}

\subsubsection{Responsibility Sealing and Accountability} Using blockchain technology, the responsibility data is sealed and cannot be changed; therefore, the authenticity and integrity of the data are guaranteed. In addition, accident accountability can be traced back in the weighted DAG.

%% file: experiment.tex
\section{IoR model in practice}

\subsection{Applied in Enterprises and Institutions}

\begin{figure}[!h]
	\centering
	\includegraphics[width=0.8\linewidth, height=0.3\textheight]{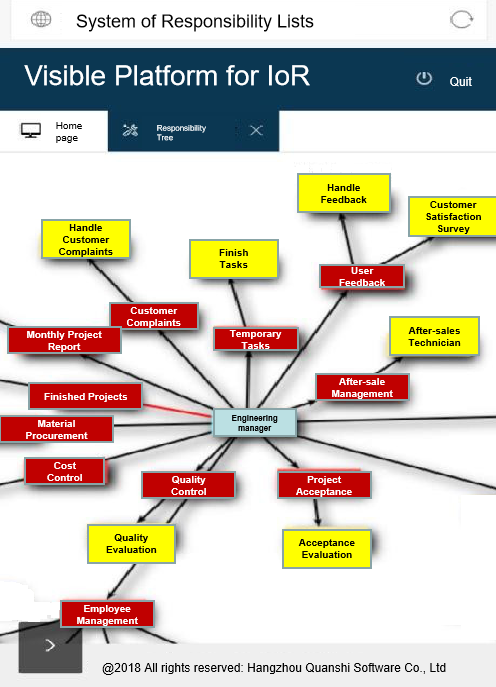}
	\vspace{0mm}
	\caption{Generated tree by responsibility lists.}
	\label{fig:figure3}
	\vspace{-6mm}
\end{figure}

\begin{figure}[!h]
	\centering
	\includegraphics[width=0.8\linewidth, height=0.16\textheight]{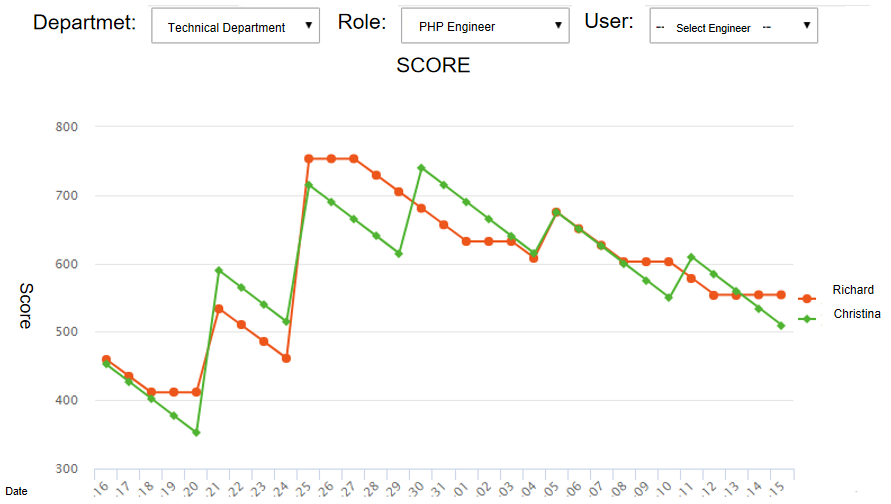}
	\vspace{-1mm}
	\caption{Responsibility score.}
	\label{fig:figure4}
	\vspace{-3mm}
\end{figure}

Figure~\ref{fig:figure3} shows the deployment of the proposed IoR model in Hangzhou Quanshi Software Co., Ltd. The corresponding tree of responsibility lists is automatically generated through a mobile application. Also, staff can submit relevant data through the app and the system can thereafter perform evaluation and scoring automatically. The system computes the personal responsibility score (rank) in the department or in the organization; as well as the personal score (status) based on the due diligence applied towards accomplishing the responsibility assigned. Figure~\ref{fig:figure4} reflects the responsibility scores. In the case of low scores, the system notifies/warns employees to complete their responsibilities.

\label{sect:experiment}

\subsection{Applied in Government Accountability}
The IoR model has been applied in government accountability with respect to the following aspects: 
\subsubsection{Safety Evaluation}
With the sealed IoR data model, safety evaluation scores are displayed on the system.
\subsubsection{Accountability Mechanism}
\begin{itemize}
    \item[-] Record accountability data in unsafe situations.
    \item[-] Design the corresponding responsibility reminder alarm system, based on responsibility score.
\end{itemize}

\subsubsection{Responsibility Ranking}
Safety level is ranked according to the responsibility score among units.

\subsubsection{Supervised Risk Level}
\begin{figure}[!ht]
	\centering
	\vspace{-2mm}
	\includegraphics[width=0.8\linewidth, height=0.15\textheight]{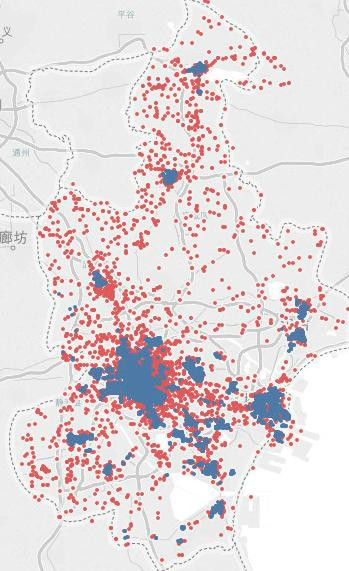}
	\caption{Responsibility status map}
	\label{fig:figure5}
	\vspace{0mm}
\end{figure}

Risk level in various units is summarized by the system so that units with decreasing responsibility scores can be immediately reminded. Risk management maturity reflects the risk management capability of an enterprise from various aspects~\cite{Sun_2019}. Figure~\ref{fig:figure5} shows risk level (decreasingly ranked by colors red, yellow, green, blue), allowing governments to have an oversight of their agencies with near real-time supervision and monitoring capabilities.

%% file: conclusion.tex
\section{Conclusion}

To address the challenging issue of allocating responsibilities in a production environment, this work proposed the notion of the Internet of Responsibilities and implemented an IoR system driven by big data and blockchain technology. The practical deployments showed an obvious increase in the responsibility scores and safety awareness in both personnel and organizations.